\documentclass[conference]{IEEEtran}
\ifCLASSINFOpdf
\usepackage[pdftex]{graphicx}
\else
\usepackage[dvipdfmx]{graphicx}
\fi
\usepackage{url}

\begin{document}
\noindent This paper is the preprint version of the paper ``Classification of URL bitstreams using bag of bytes'' published by IEEE with DOI: 10.1109/ICIN.2018.8401597.\\

\noindent\copyright{} 2021 IEEE. Personal use of this material is permitted. Permission from IEEE must be obtained for all other uses, in any current or future media, including reprinting/republishing this material for advertising or promotional purposes, creating new collective works, for resale or redistribution to servers or lists, or reuse of any copyrighted component of this work in other works.\\
\newpage

%
\title{Classification of URL bitstreams\\ using Bag of Bytes}



%
\author{\IEEEauthorblockN{
    Keiichi Shima\IEEEauthorrefmark{1},
    Daisuke Miyamoto\IEEEauthorrefmark{2},
    Hiroshi Abe\IEEEauthorrefmark{1},
    Tomohiro Ishihara\IEEEauthorrefmark{3},\\
    Kazuya Okada\IEEEauthorrefmark{3},
    Yuji Sekiya\IEEEauthorrefmark{3},
    Hirochika Asai\IEEEauthorrefmark{4} and
    Yusuke Doi\IEEEauthorrefmark{4}
  }
  \IEEEauthorblockA{\IEEEauthorrefmark{1}
    Internet Initiative Japan Inc.,
    2-10-1 Fujimi, Chiyoda-ku, Tokyo 102-0071, Japan\\
    Email: keiichi@iijlab.net and abe@iij.ad.jp}
  \IEEEauthorblockA{\IEEEauthorrefmark{2}
    Nara Advanced Institute of Science and Technology,
    8916-5 Takayama-cho, Ikoma, Nara 630-0192, Japan\\
    Email: daisu-m@is.naist.jp}
  \IEEEauthorblockA{\IEEEauthorrefmark{3}
    The University of Tokyo,
    7-3-1 Hongo, Bunkyo-ku, Tokyo 113-8654, Japan\\
    Email: sho@c.u-tokyo.ac.jp,
    okada@ecc.u-tokyo.ac.jp and
    sekiya@nc.u-tokyo.ac.jp}
  \IEEEauthorblockA{\IEEEauthorrefmark{4}
    Preferred Networks, Inc.,
    1-6-1 Otemachi, Chiyoda-ku, Tokyo 100-0004, Japan\\
    Email: asai@preferred.jp
    and doi@preferred.jp}
}


\maketitle

\begin{abstract}
  Protecting users from accessing malicious web sites is one of the
  important management tasks for network operators. There are many
  open-source and commercial products to control web sites
  users can access.  The most traditional approach is
  blacklist-based filtering.  This mechanism is simple but not
  scalable, though there are some enhanced approaches utilizing
  fuzzy matching technologies.  Other approaches try to use machine
  learning (ML) techniques by extracting features from URL strings.
  This approach can cover a wider area of Internet web sites, but
  finding good features requires deep knowledge of trends of web site
  design.  Recently, another approach using deep learning (DL) has
  appeared.  The DL approach will help to extract features
  automatically by investigating a lot of existing sample data.  Using
  this technique, we can build a flexible filtering decision module by
  keep teaching the neural network module about
  recent trends, without any specific expert knowledge of the URL domain.
  In this paper, we
  apply a mechanical approach to generate feature vectors
  from URL strings.  We implemented our approach and tested with
  realistic URL access history data taken from a research organization
  and data from the famous archive site of phishing site information,
  \textit{PhishTank.com}.  Our approach achieved 2\textasciitilde 3\% better
  accuracy compared to the existing DL-based approach.
\end{abstract}


%
\IEEEpeerreviewmaketitle

\section{Introduction}

As the Internet grows and becomes more stable, the importance of its
networking function as a social infrastructure becomes greater and
greater.  Although abuse was observed from
the beginning of the Internet, as the Internet is being relied on by
business activities etc.,
business-oriented attacks have increased.  From the
viewpoint of network management, we need to consider several
different kinds of threats to protect both network operational
stability and the users who are connected and using the network.  We need to protect against (D)DoS attacks, unintended network accesses, virus infection, and so on.

Phishing is one such threat to which network operators must pay
attention.  An attacker tries to make users access faked sites that
look similar to existing web service pages, typically banking
sites or shopping sites.  In such faked pages, the attacker provides a
faked login screen to steal account numbers and passwords of victims.
A document\cite{apwg2017-report} published by Anti-Phishing Working
Group\footnote{\url{https://antiphishing.org/}} revealed that more
than 1.2 million phishing attacks were reported in 4Q 2016 which
is 65\% larger than 4Q 2015.  Because the document covered only
reported attacks, we can guess there was more hidden or
unnoticed attacks in reality.

Network operators need to protect their users from accessing such
malicious sites.  If you are an operator of an ISP or a similar kind
of service aggregator, then you need to take care of your customer's
network service operations too.  Since the number of phishing sites is
huge and growing (100 of thousands of unique sites are reported in the
document\cite{apwg2017-report} and they keep changing their site
locations), we need an automated and adaptive approach to defend
customers from such activities.

The rest of the paper is structured as follows. We discuss related
work in Section~\ref{sec:related-work}.  Our proposed idea to form a
unique URL feature vector is introduced in Section~\ref{sec:bob-idea}.
The neural network topology used to classify URLs in our experiments
is disclosed in Section~\ref{sec:neural-network-topology}.  The
datasets and evaluation results are explained in
Section~\ref{sec:datasets} and \ref{sec:evaluation}.  We
conclude the achievementss and future directions of this work in
Section\ref{sec:conclusion}.

\section{Related Work}
\label{sec:related-work}

Garera \textit{et al.} proposed a framework to detect phishing sites
using machine learning technique in \cite{garera2007-phishing}.  They
defined several features which they thought important to classify
benign URLs and malicious URLs.  Examples of the features are a page
rank value, a type of domain names (e.g. whether the host part
consists of IP address numbers or a hostname,
whether the number of sub-domains are too large, etc), and so
on.  Ma \textit{et al.} proposed \cite{ma2009-beyond-blacklist} a URL
classification method using very high dimensional feature vectors
generated from lexical features of URL strings and host name features
such as reputation.  Prakash \textit{et al.} proposed a new blacklist
generation mechanism in \cite{prakash2010-phishnet}.  Their idea is
that attackers often use similar URLs to those they were using before and use slightly modified strings in new phishing attacks.
They proposed some rules to generate potential blacklist URLs to
protect future phishing attacks.

So far, the previous work is based on string analysis and external
information (such as page rank) to define feature vectors of URLs.
Since they are based on expert knowledge studied when they were
working on the classification issue, if the assumptions of the
knowledge changes then the mechanism may lose accuracy.

The other approach is to inspect the contents of the target URLs.
Zhang \textit{et al.} proposed a phishing site detection mechanism by
looking into the contents of the target sites\cite{zhang2007-cantina}.
This approach apparently works more precisely; however accessing the
target sites causes different issues, such as network traffic
overhead, possible risks to access (possibly) malicious contents.

Saxe \textit{et al.} proposed a deep neural networking-based approach
named eXpose\cite{saxe2017-expose} to classify URLs, file names, and
registry keys.  In contrast to the past works, they tried not to
define any feature vectors using expert knowledge.  Instead, they
tried to offload the work to extract feature vectors to their deep
neural network model.  They expand each character of a URL string
using the embedding technique and apply four different sizes of
convolutional neural networking layers to extract URL features.  In
their paper, their approach achieved better accuracy than an expert
knowledge based feature extraction approach and an N-gram based
feature extraction approach. They imply better
accuracy may be achieved by using different URL vectorization methods
and/or different neural network topologies.  In this paper, we follow
this approach and try to achieve simpler and better result.

\section{Bag of Bytes and URL Vector}
\label{sec:bob-idea}

We tried not to use any expert knowledge of a specific domain, in this
case, knowledge of URL structures.  In that sense, the direction of
our approach is same as eXpose\cite{saxe2017-expose} took.  In
eXpose, they used the last 200 characters of a URL string and converted it
into a $200\times32$-dimensional vector using an embedding
technique. The reason why they cut the string to 200 characters is
that 95\% of URL strings are shorter than 200 characters in their
investigation.  In our approach, we also try to convert URL strings
mechanically but use all the characters included in the URL strings.

\begin{figure}[!t]
  \centering
  \includegraphics[width=0.9\columnwidth]{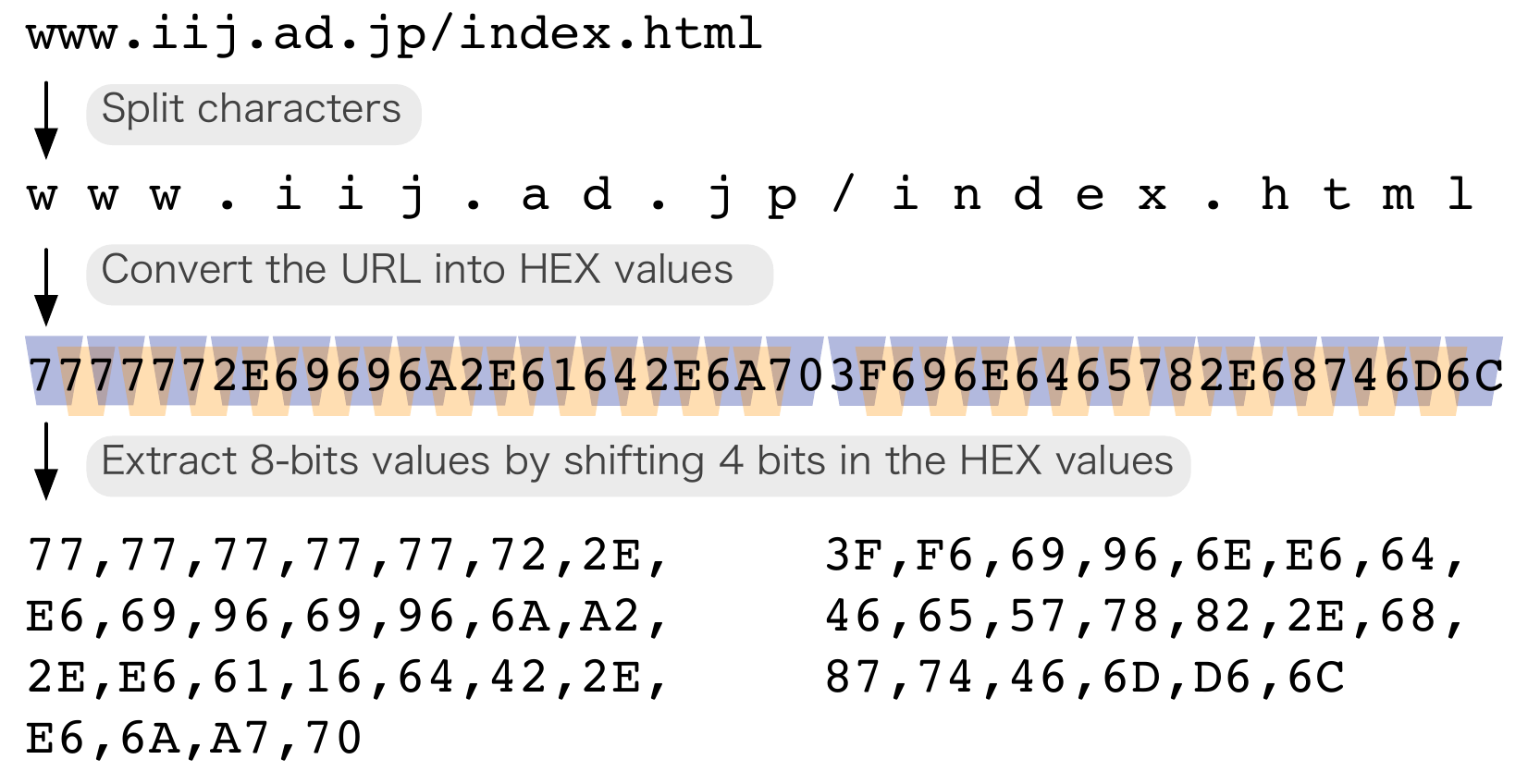}
  \caption{The procedure to extract byte values from a URL string}
  \label{fig:extract-byte-values}
\end{figure}

Fig.~\ref{fig:extract-byte-values} shows how a URL string is
converted.  The basic idea is to count byte values included in the URL
string.  The point is that we extract byte values not only from each
character but also overlapping parts of neighboring characters by
shifting 4-bits when extracting byte values.  By using intersecting
bits between two neighboring cahracters, we can embed combination
information of two characters appearing sequentially.
After extracting the byte values, we count how many times each
value appears in the original URL string.  We perform the extraction
operation for both the host part and the path part separately, and
achieve a 512-dimensional vector that represents the original URL.
Finally, each vector is normalized and we treat the final vector value
as a URL vector.

\section{Neural Network Topology}
\label{sec:neural-network-topology}

\begin{figure}[!t]
  \centering
  \includegraphics[width=0.8\columnwidth]{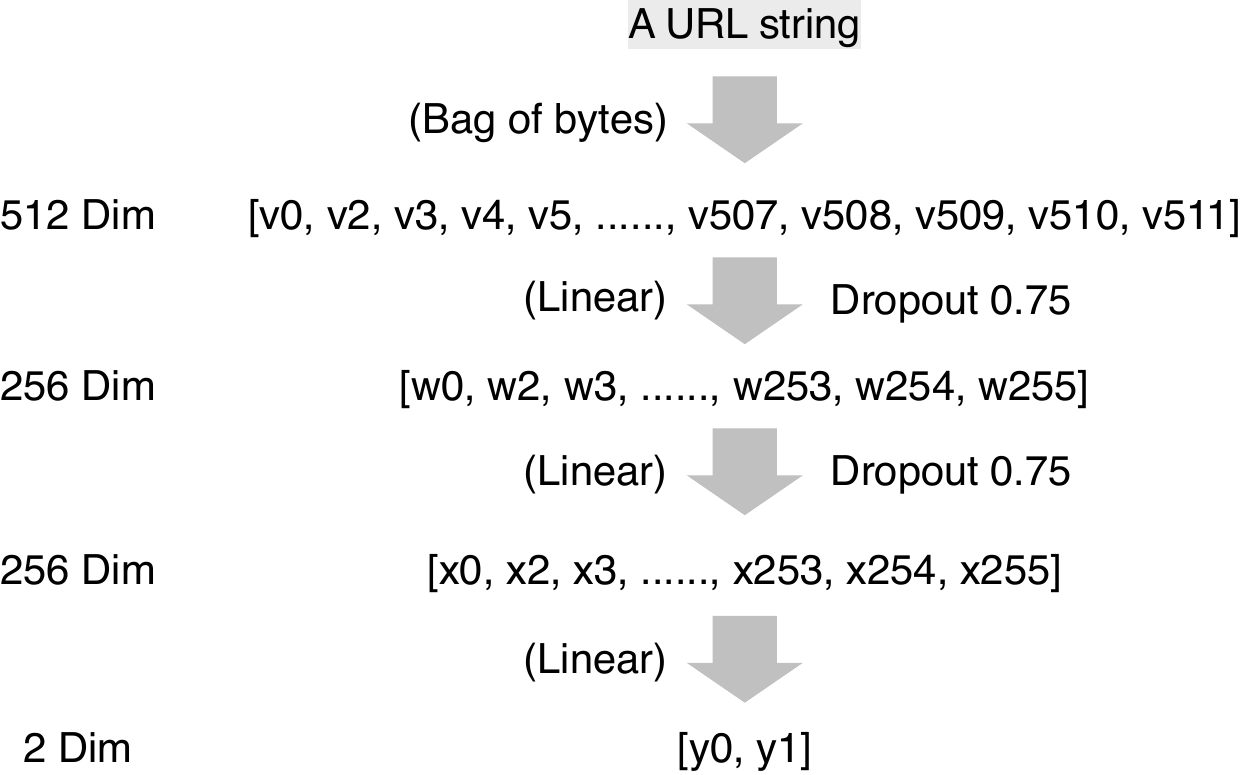}
  \caption{The neural network model for URL classification}
  \label{fig:dnn-model}
\end{figure}

The neural network topology we used for classification is not complex.
The topology is a basically multi-layer linear (or sometimes called a
`dense') topology.  Fig.~\ref{fig:dnn-model} depicts the neural
network topology for URL classification.

The first layer is a 512-dimensional array of nodes which receives the
vector value of a URL, the first half comes from the host part, the
latter part comes from the path part.  The value will be passed to a
256-dimension array twice using linear connections.  To suppress
over-fitting, links are dropped out with the ratio of 0.75.  The
dropout ratio value is decided based on our experience of several
trials of training.  Finally, the 256-dimension array will be mapped to
two classes that indicate whether the input URL is a benign URL or a
malicious URL.

\section{Datasets}
\label{sec:datasets}

The datasets we used to evaluate the proposed model consist of two
sources.  One is a set of URLs retrieved from the
\textit{PhishTank.com}\footnote{\url{https://www.phishtank.com/}}.  The site
provides information on phishing sites based on the reports submitted
by users. The submitted URLs are manually checked to see if the URL is a
real phishing site; if so, it is marked as active.  PhishTank.com
distributes latest active phishing site list which can be used for
access filtering.  We retrieved the recorded set of phishing URLs
and used them as a blacklist.

For a whitelist, we collected access log data of a research
organization.  A difficult point is that it is impossible to prepare
a pure white URL access list from a real access log.  There is always
the possibility that the list we collected includes malicious URLs.  To
minimize the effect, we try to clean up the access list by removing all
the URL entries listed in the phishing site list retrieved from
PhishTank.com.

Table~\ref{tab:datasets} shows the datasets we prepared for
training.  The number of entries in each dataset is also shown in
the table.

The main target is the URL access list captured at an anonymous
research organization \textit{X} on 2017-04-25.  The access list
contains more than 142 million
entries.  The list contains not only benign URLs but also phishing
site URLs.  We tried to clean the list with the phishing site URL
data reported at PhishTank.com from 2017-04-24 to 2017-10-03.  Using
the blacklist data, including future entries beyond the target data,
will help to remove some of URLs that had not been found at the
day of 2017-04-25 and make the white URL a bit whiter.

We prepared a balanced dataset to fit the neural network for both
malicious URL features and benign URL features evenly.  Since the
number of the white URSs was larger than that of the black URLs, we
first picked 10,000 entries of the URL access log from each hour,
i.e. 240,000 entries, and randomly selected the 26,722 entries from
the list which was the same size as Blacklist 1.

\begin{table}[!t]
  \renewcommand{\arraystretch}{1.3}
  \caption{URL datasets for training}
  \label{tab:datasets}
  \centering
  \begin{tabular}{|l||p{0.6\columnwidth}|r|}
    \hline
    Type & Content & Count \\
    \hline
    \hline
    Blacklist 1
    & Phishing site URLs reported at PhishTank.com before
    \underline{2017-04-25}.
    This list is used as a blacklist for learning and testing
    in conjunction with the Whitelist 1.
    & 26,722\\
    \hline
    Blacklist 2
    & Phishing site URLs reported at PhishTank.com before
    \underline{2017-10-03}.
    This list is used to cleanse the target access log captured at the
    anonymous research organization X.
    & 68,172\\
    \hline
    Whitelist 1
    & A sampled list of URL access log captured at the anonymous
    research organization X on \underline{2017-04-25}
    excluding the entries listed in
    the Blacklist 2.  This list is used for learning and testing
    in conjunction with the Blacklist 1.
    & 26,722\\
    \hline
  \end{tabular}
\end{table}

\section{Evaluation}
\label{sec:evaluation}

We implemented our idea described in
section~\ref{sec:neural-network-topology} using
\textit{Chainer}\footnote{\url{https://chainer.org}}.  The code of the
model is shown in Fig.~\ref{fig:dnn-model-code}.

\begin{figure}[!t]
  \centering
  \footnotesize
\begin{verbatim}
from chainer import Chain
import chainer.functions as F
import chainer.links as L
class Model(Chain):
    def __init__(self):
        super(Model, self).__init__()
        with self.init_scope():
            self.l1 = L.Linear(None, 256)
            self.l2 = L.Linear(None, 256)
            self.l3 = L.Linear(None, 2)
    def __call__(self, x):
        h1 = F.dropout(F.relu(self.l1(x)),
                       ratio=0.75)
        h2 = F.dropout(F.relu(self.l2(h1)),
                       ratio=0.75)
        y = self.l3(h2)
        return y
\end{verbatim}
  \caption{The code fragment that implements our proposed neural
    network model using Chainer}
  \label{fig:dnn-model-code}
\end{figure}

We used the datasets shown in Table~\ref{tab:datasets}.  The URL
entries included in the Blacklist 1 and the Whitelist 1 are mixed
and randomly shuffled.  The ratio of training and validating is 80\% and
20\%.  The mini-batch size is set to 100, and the number of epochs is
20.

We tested our neural network model using three different optimizers,
\textit{Adam}, \textit{AdaDelta}, and \textit{SGD}.  Among them, Adam
was the best optimizer with an accuracy of 94.18\%.

As mentioned above, eXpose\cite{saxe2017-expose} tried to
classify URLs using a convolutional neural network.  Unfortunately,
while they described their neural network model, they didn't
provide their code and dataset used in their evaluation.  In their
paper, they said they achieved more than 99.9\% accuracy.  To
compare their approach to ours, we implemented their neural network
model using Chainer and evaluated it with the same dataset we used for
our cases.  The results are also shown in the same table.  Although
the result using SGD as an optimizer was a bit low; however, with the
other optimizers, their approach achieved almost same but a bit lower
accuracy than ours.  We also measured the time consumed for training
because they are using more complex neural networking topology.  Their
model requires four times more training time than ours.

\begin{table}[!t]
  \renewcommand{\arraystretch}{1.3}
  \caption{Results of accuracy and training time using Whitelist 1 and
    Blacklist 1 in Table~\ref{tab:datasets}}
  \label{tab:results}
  \centering
  \begin{tabular}{|c||c|r|r|}
    \hline
    & Optimizer & Accuracy (\%) & Training time (s)\\
    \hline
    \hline
    Our method & Adam & 94.18 & 32\\
    \hline
    -- & AdaDelta & 93.54 & 31\\
    \hline
    -- & SGD & 88.29 & 31\\
    \hline
    \hline
    eXpose\cite{saxe2017-expose} & Adam & 90.52 & 119\\
    \hline
    -- & AdaDelta & 91.31 & 119\\
    \hline
    -- & SGD & 77.99 & 116\\
    \hline
  \end{tabular}
\end{table}

\begin{figure*}[!t]
  \centering
  \includegraphics[width=0.8\textwidth]{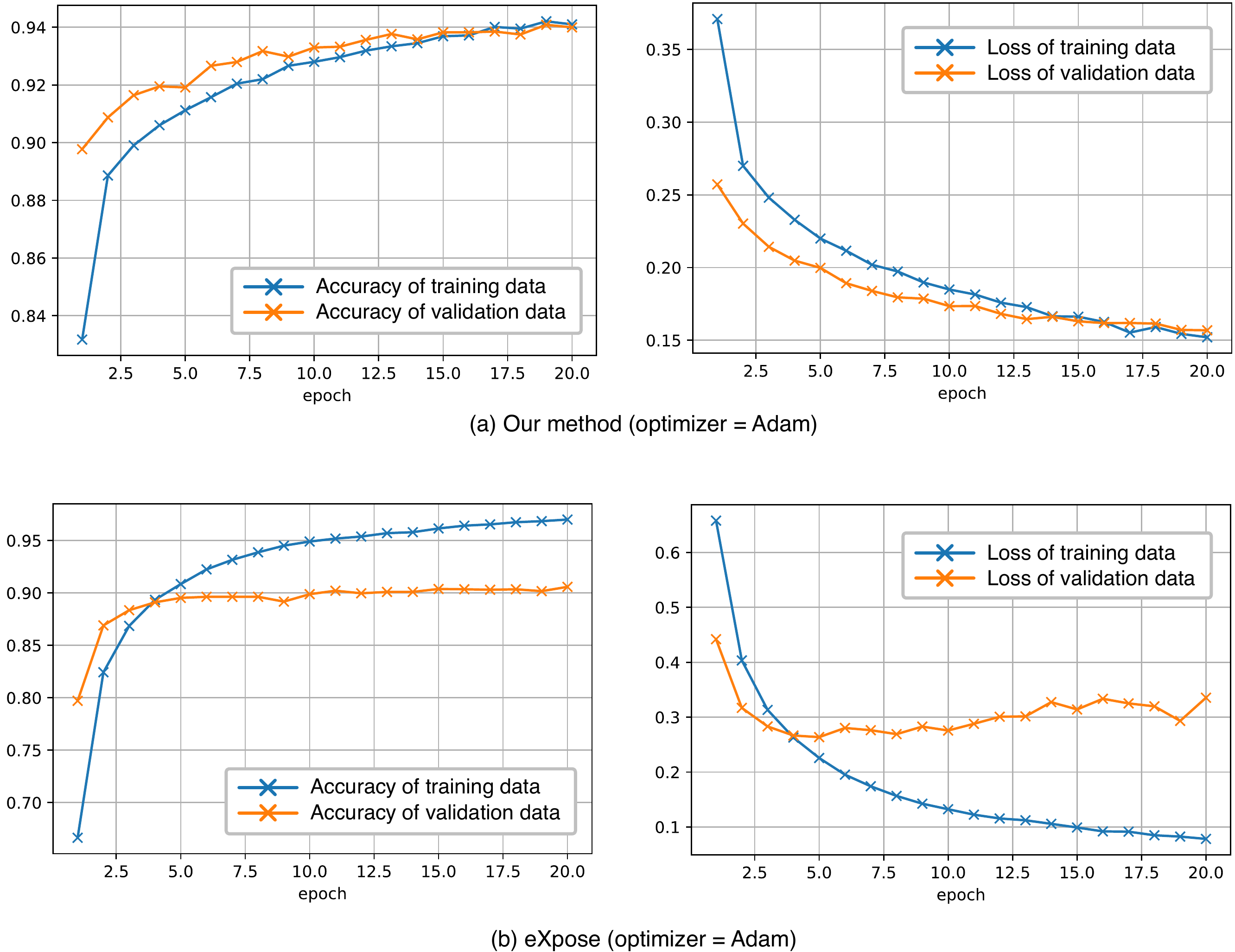}
  \caption{Learning curves of our method (a) and eXpose (b): The blue
    lines indicate results of training data and the orange lines
    indicate results of validation data of each epoch while training
    each model.}
  \label{fig:learning-curves}
\end{figure*}

Fig.~\ref{fig:learning-curves} shows the learning curves (accuracy
and loss values at each epoch).  eXpose quickly converged to the
stable state compared to our method; although the final accuracy is
lower than ours using our datasets.  When looking at the loss values,
eXpose looks to be over-fitting when the count of epochs increase.  Our
proposal uses a dropout ratio of 0.75 between neural network layers
to suppress over-fitting,
while eXpose uses 0.5 as is specified in the eXpose paper.
The larger dropout value may contribute less over-fitting in
the eXpose case.

\begin{table}[!t]
  \renewcommand{\arraystretch}{1.3}
  \caption{URL datasets for predicting}
  \label{tab:datasets2}
  \centering
  \begin{tabular}{|l||p{0.6\columnwidth}|r|}
    \hline
    Type & Content & Count \\
    \hline
    \hline
    Blacklist 3
    & Phishing site URLs reported at PhishTank.com before
    \underline{2017-05-25}.
    This list is used as a black list for learning and testing
    in conjunction with the white list.
    & 39,776\\
    \hline
    Whitelist 2
    & A sampled list of URL access log captured at the anonymous
    research organization X on \underline{2017-05-25} excluding
    the entries listed in
    the Blacklist 2.  This list is used for test the neural network
    model trained with the Blacklist 1 and Whitelist 1.
    & 39,776\\
    \hline
  \end{tabular}
\end{table}

\begin{figure}[!t]
  \centering
  \includegraphics[width=0.9\columnwidth]{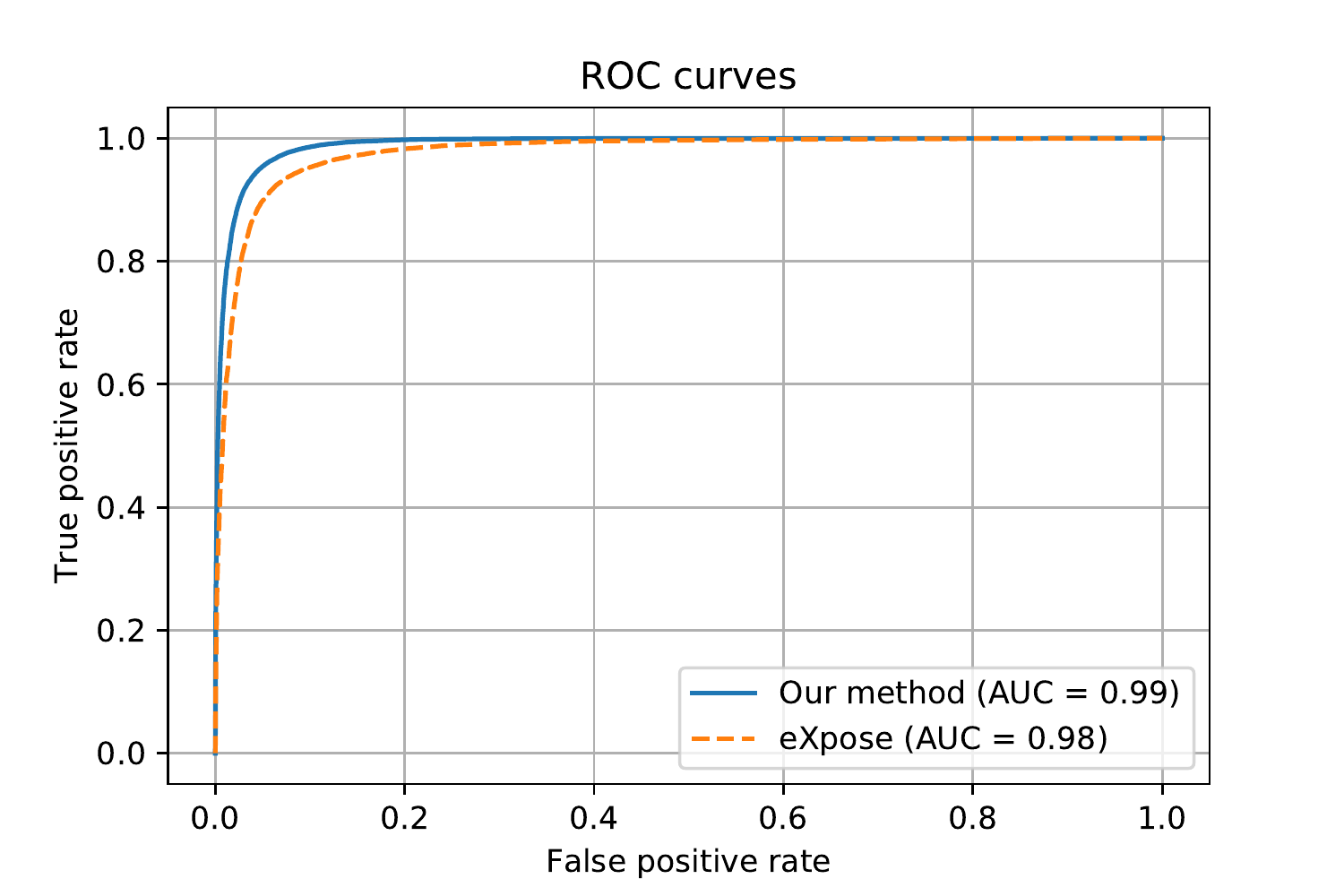}
  \caption{ROC curves and AUC values measured with the prediction
    datasets as shown in Table~\ref{tab:datasets2} using our model and
    eXpose model}
  \label{fig:roc-curves}
\end{figure}

We tried to apply the neural network model trained with the dataset of
Blacklist 1 and Whitelist 1 on a different dataset containing data
captured later than the training data as shown in
Table~\ref{tab:datasets2}.  The evaluation results are shown in
Table~\ref{tab:prediction-results}.  Our method achieved 95.17\% of
accuracy with 0.9525 of F-measure score.  We tried to predict the same
dataset with the eXpose model trained with the same trainer dataset
too.  The results are also shown in
Table~\ref{tab:prediction-results}.  eXpose achieved good but slightly
lower score than our method.  The Receiver Operating Characteristic
(ROC) curves and Area Under the Curve (AUC) values are shown in
Fig.~\ref{fig:roc-curves}.

\begin{table}[!t]
  \renewcommand{\arraystretch}{1.3}
  \caption{Prediction results of the dataset shown in
    Table~\ref{tab:datasets2} using the trained neural network model
    with the dataset shown in Table~\ref{tab:datasets}}
  \label{tab:prediction-results}
  \centering
  \begin{tabular}{|l||r|r|r|r|}
    \hline
    & Accuracy (\%) & Precision (\%) & Recall (\%) & F-measure \\ \hline
    \hline
    Our method & 95.17\% & 93.76\% & 96.78\% & 0.9525 \\ \hline
    eXpose & 92.99\% & 93.00\% & 92.99\% & 0.9299 \\ \hline
  \end{tabular}
\end{table}

It is difficult to say if our neural network model is applicable to a
specific real operation or not given the results shown only in this
paper.  The accuracy may change depending on the target environment.
The eXpose paper showed their classification accuracy was more than
99.9\%; however as we show, if we change the dataset the result
changes.  For the same reason, if the data sources are changed, our
mechanism may produce lower accuracy than in this paper.

One of the issues when working on this type of research is that using
a generalized dataset is really difficult.  The approaches sometimes
optimized to the target datasets, which are the only datasets that the
authors can access in many cases.  In the image recognition field,
researchers have several common datasets such as
MNIST\footnote{\url{http://yann.lecun.com/exdb/mnist/}} that can be
used to evaluate each researchers' proposal with a same baseline.
Probably we need to start the effort to build shared datasets for
network data too.

\section{Conclusion}
\label{sec:conclusion}

In this paper, we proposed a new neural network model for classifying
URLs into benign and phishing.  The learning overhead of the proposed
network is light because the topology consists of just three layers of linear
networks.  Shorter learning time make it possible to try with many
different kinds of data to optimize the neural network topology.
Since the accuracy is sometimes affected by the quality of training
data, more trials may result in more suitable networks.

The debatable point is that it is not possible to prove which method
is better or best.  In this area, analyzing network log data using
neural network technologies, we lack a common dataset to compare
approaches.  In this paper, we could achieve better performance than the
past work, however, it might be worse if we used other datasets.
Thus, all the researchers in this community need to start building a
dataset to share among researchers working on network log analysis
using machine/deep learning approaches to improve the technologies.

The datasets we used in this paper were taken from the famous
phishing URL archive site PhishTank.com and from an anonymous research
organization's network.  With these realistic datasets, our model
resulted in 94.18\% accuracy in the training/validation phase and
95.17\% when classifying newer dataset then the training phase, which
was higher than the
previously proposed neural network model published in
\cite{saxe2017-expose} which uses more complex neural network
topology.

\section*{Acknowledgment}
This work was supported by JST CREST
Grant Number JPMJCR1783,
Japan.

\bibliographystyle{IEEEtran}
\bibliography{ni2018-url-clf}

\end{document}